\documentclass{aa}  

\usepackage{graphicx}
\usepackage{natbib}
\usepackage[utf8]{inputenc}
\usepackage{natbib}
\usepackage[mathscr]{euscript}
\usepackage{lscape}
\usepackage{float}
\usepackage{nonfloat}
\usepackage{rotating}
\usepackage{dcolumn}
\newcommand\hill{Hill et al. (in prep.)}
\newcommand\Teff{$T_\textsl{eff}$}
\newcommand\logg{$\log g$}
\newcommand\vt{$v_t$}

\begin{document}

   \title{Sulphur in the Sculptor dwarf spheroidal galaxy\thanks{Based on observations made with ESO/VLT/FLAMES at the La Silla Paranal observatory under program ID 089.B-0304(B)}$^,$\thanks{Tables \ref{tab:main}, \ref{tab:abulines}, and \ref{tab:nlte}. are available online, but examples of these tables are given at the end of this document.}}

   \subtitle{Including NLTE corrections}

   \author{\'{A}. Sk\'{u}lad\'{o}ttir
                \inst{1}
        \and
        S. M. Andrievsky \inst{2,3}                   
        \and
        E. Tolstoy\inst{1}
        \and
        V. Hill \inst{4}
        \and
                S. Salvadori\inst{1}
        \and            
        S. A. Korotin\inst{2}         
                \and
        M. Pettini \inst{5}  
          }
          
   \institute{
              Kapteyn Astronomical Institute, University of Groningen, PO Box 800, 9700AV Groningen, the Netherlands\\
              \email{asa@astro.rug.nl}
                \and
                        Department of Astronomy and Astronomical Observatory, Odessa National University, Ukraine, and Isaac Newton Institute of Chile, Odessa branch, Shevchenko Park, 65014, Odessa, Ukraine            
                \and    
                        GEPI, Observatoire de Paris-Meudon, CNRS, Universite Paris Diderot, 92125 
Meudon Cedex, France            
                \and
             Laboratoire Lagrange, Universit\'{e} de Nice Sophia Antipolis, CNRS, Observatoire de la C\^{o}te d’Azur, CS34229, 06304 Nice Cedex 4, France
                \and
                Institute of Astronomy, Madingley Road, Cambridge CB3 0HA, England
                }

\abstract{
In Galactic halo stars, sulphur has been shown to behave like other $\alpha$-elements, but until now, no comprehensive studies have been done on this element in stars of other galaxies. Here, we use high-resolution ESO VLT/FLAMES/GIRAFFE spectra to determine sulphur abundances for 85 stars in the Sculptor dwarf spheroidal galaxy, covering the metallicity range $-2.5\leq \text{[Fe/H]} \leq-0.8$. The abundances are derived from the S~I triplet at 9213, 9228, and 9238~\AA. These lines have been shown to be sensitive to departure from local thermodynamic equilibrium, i.e. NLTE effects. Therefore, we present new NLTE corrections for a grid of stellar parameters covering those of the target stars. The NLTE-corrected sulphur abundances in Sculptor show the same behaviour as other $\alpha$-elements in that galaxy (such as Mg, Si, and Ca). At lower metallicities ($\text{[Fe/H]}\lesssim-2$) the abundances are consistent with a plateau at $\text{[S/Fe]}\approx+0.16$, similar to what is observed in the Galactic halo, $\text{[S/Fe]}\approx+0.2$. With increasing [Fe/H], the  [S/Fe] ratio declines, reaching negative values at $\text{[Fe/H]}\gtrsim-1.5$. The sample also shows an increase in [S/Mg] with [Fe/H], most probably because of enrichment from Type Ia supernovae.

}

   \keywords{Stars: abundances --
                                Galaxies: dwarf galaxies --
                                Galaxies: individual (Sculptor dwarf spheroidal) --
                                Galaxies: abundances --
                                Galaxies: evolution
               }

   \maketitle

%
\section{Introduction}

The chemical abundances of the photospheres of stars reveal the composition of their birth environment; studying stars of different ages, therefore, gives information about the chemical enrichment history of the galaxy in which they dwell. The element sulphur is one of the least studied $\alpha$-elements in stars of the Galactic halo and the dwarf galaxies surrounding the Milky Way. This is unfortunate, since sulphur is not believed to be heavily bound onto interstellar dust \citep{Jenkins2009}, and the abundances of sulphur in stars can therefore be directly compared to measurements in interstellar medium, such as damped Lyman-$\alpha$ absorption (DLA) systems \citep{Nissen2007}.

In general, most stellar abundance analysis is done by assuming local thermodynamic equilibrium (LTE). Depending on the lines in question, this can be a reasonable assumption, but sulphur lines have been shown to be sensitive to effects caused by deviation from LTE, i.e.~NLTE corrections are needed \citep{Takeda2005,Korotin2008}. Furthermore, the S~I lines of multiplet~1, which is most commonly used to measure sulphur in metal-poor stars, sit in a rather challenging wavelength region, $ 9210$-$9240$~\AA, which is covered with telluric absorption lines. 

Two recent surveys of sulphur in the Galactic halo, by \citet{Nissen2007} and \citet{Spite2011}, measure sulphur abundances from multiplet~1, properly treating the NLTE effects on the measured lines. The observed trend of [S/Fe] with [Fe/H] in these studies\footnote{$\text{[S/Fe]}=\log_{10}(N_{\text{S}}/N_\text{Fe})_\star - \log_{10}(N_{\text{S}}/N_\text{Fe})_\odot$} is very similar to other $\alpha$-elements (such as Mg, Si, and Ca). The main production sites of the $\alpha$-elements are core-collapse Type~II Supernovae (SNe). Early in the star formation history of any galaxy, Type II SNe are believed to be the main contributors of metals, so the early interstellar medium (ISM) holds the imprint of their yields, and ancient stars typically show an enhancement in [$\alpha$/Fe]. This is consistent with the sulphur results from \citet{Nissen2007} and \citet{Spite2011}, which show an approximately constant value of $\text{[S/Fe]}~\approx~+0.2$ at low metallicities. 

Other studies of sulphur have been done for metal-poor stars, using multiplet 3 at 10455-10459~\AA, which is relatively free from telluric contamination. These studies are also consistent with a plateu of enhanced [S/Fe] values at low metallicities, both in the Galactic halo \citep{Caffau2010,Jonsson2011,TakedaTakada-Hidai2012} and also in globular clusters \citep{Kacharov2015}.

\begin{table}
\caption{Log of the VLT/GIRAFFE service mode observations.}
\label{tab:obs}
\centering
\begin{tabular}{c c c c c}
\hline\hline
Date    &       Plate   &       Exp.time        &       Airmass &       Seeing  \\
        &               &        (min)  &       (average)       &       (arcsec)        \\
\hline
2012-Jul-15      &       MED1            &       51.25   &       1.078   &       1.46    \\
2012-Jul-15      &       MED1            &       51.25   &       1.026   &       1.43    \\
2012-Jul-27      &       MED2            &       51.25   &       1.028   &       1.53    \\
2012-Jul-27      &       MED2            &       51.25   &       1.020   &       1.59    \\
2012-Jul-27      &       MED2            &       51.25   &       1.100   &       1.48    \\
2012-Aug-21      &       MED2            &       51.25   &       1.025   &       1.58    \\
2012-Aug-21      &       MED1            &       51.25   &       1.027   &       1.41    \\
2012-Aug-21      &       MED2            &       51.25   &       1.104   &       1.44    \\
2012-Aug-21      &       MED2            &       51.25   &       1.211   &       1.43    \\
2012-Aug-21      &       MED2            &       51.25   &       1.082   &       1.47    \\
2012-Aug-23      &       MED2            &       51.25   &       1.242   &       1.56    \\
2012-Aug-23      &       MED2            &       51.25   &       1.107   &       1.62    \\
\hline
\end{tabular}
\end{table}

At higher metallicities, $\text{[Fe/H]}\gtrsim-1$, there is  a decline in [S/Fe] with [Fe/H] in Galactic disc stars \citep{Clegg1981,Francois1987,Francois1988,Chen2002}. This mirrors the trends of other $\alpha$-elements, and is believed to be caused by the yields from Type~Ia SNe, which start to explode about 1-2~Gyr after the first Type~II SNe  (e.g. \citealt{MatteucciGreggio1986,WyseGilmore1988,deBoer2012}) and produce large amounts of the iron-peak elements compared to the $\alpha$-elements \citep{Iwamoto1999}, leading to a decrease in the [$\alpha$/Fe] ratio.

Until now, no surveys of sulphur have been made of individual stars in galaxies other than the Milky Way. Various $\alpha$-elements in stars in dwarf spheroidal (dSph) galaxies have been shown to have a different behaviour from the Milky Way (e.g. \citealt{Shetrone2001,Shetrone2003,Tolstoy2003,Tolstoy2009,Kirby2011}). At the lowest metallicities they show a high abundance in [$\alpha$/Fe] similar to the Galactic halo, but the decline due to Type~Ia~SNe starts at a lower [Fe/H] than in the Milky Way, the exact value depending on the galaxy. At the highest metallicities in these galaxies, these ratios reach negative values, $\text{[$\alpha$/Fe]}<0$, while in the Galactic disc, these elements settle around $\text{[$\alpha$/Fe]}\approx0$, at $\text{[Fe/H]}=0$.

Sculptor is a well-studied system, with an absolute magnitude of $M_V\approx-11.2$ and a distance of $86\pm 5$~kpc \citep{Pietrzynski2008}. It is at high Galactic latitude ($b=-83^\circ$) with a systemic velocity of $v_{\textsl{hel}}~=~+110.6~\pm~0.5$~km/s. The contamination by foreground Galactic stars is not significant, and most of it can easily be distinguished by velocity (e.g. \citealt{Battaglia2008b}). The galaxy is dominated by an old stellar population (>10~Gyr), and does not appear to have experienced any episodes of star formation over the last $\sim$6~Gyr \citep{deBoer2012}.

Large spectroscopic surveys of individual stars have been carried out in the central field of Sculptor. Abundances have been measured for Fe, Mg, Si, Ca, and Ti with intermediate-resolution (IMR) spectroscopy \citep{Kirby2009} and high-resolution (HR) spectroscopy for $\sim$100 stars by the DART survey (\citealt{Tolstoy2009}; Hill et al. in prep.). These works have shown that the `knee' in Sculptor, where the contribution from Type Ia SN leads to a decline in [$\alpha$/Fe], happens around $\text{[Fe/H]}\sim-1.8$.

Similar to S, the iron-peak element Zn is volatile and not heavily depleted onto dust in interstellar gas \citep{Jenkins2009}. The [S/Zn] ratios in stars can therefore be directly compared to these ratios measured in DLA systems \citep{Nissen2007}. In this paper we present the results for sulphur in 85 stars in Sculptor, and in an upcoming paper (Sk\'{u}lad\'{o}ttir et al. in prep.) we will present Zn abundances for the same target stars and use these results to make a detailed comparison with DLA systems and the Galactic halo.

%
\section{Observations and data reduction}

\subsection{Observations}
From the DART survey \citep{Tolstoy2006}, detailed abundance measurements are known for $\sim$100 stars spread over a $25^\prime$ diameter field of view in the Sculptor dSph (\citealt{Tolstoy2009}; Hill et al. in prep.). Because of the distance to Sculptor, only the brightest stars are available for HR spectroscopy. This sample, therefore, consists of relatively cool red giant branch (RGB) stars, with $T_\textsl{eff}~\lesssim~4700$~K. For an overlapping sample of 86 stars in Sculptor, ESO VLT/GIRAFFE spectroscopy was carried out to measure  sulphur. 

GIRAFFE \citep{Pasquini2002}  is a medium- to high-resolution spectrograph, with settings covering the entire visible range and near-infrared (IR) range, 3700-9500~\AA. The GIRAFFE/MEDUSA fibres allow up to $\sim$130 separate objects (including sky fibres) to be observed in one go. Two separate sets of MEDUSA fibres exists, one per positioner plate, and each fibre has an aperture of 1.2 arcsec on the sky.

The observations presented here were taken in service mode in July and August of 2012, all using the HR22B grating, which covers the wavelength range 8960-9420~\AA, with resolution R$\sim$19,000. The observational details are listed in Table~\ref{tab:obs}.

One of the target stars, ET0097, showed very strong CN molecular lines. A higher resolution, follow-up spectrum for that star was taken with ESO VLT/UVES, covering a longer wavelength range. The stellar parameters and LTE abundances for sulphur presented here for this star are taken from these extended data \citep{Asa2015}.

\subsection{Data reduction}

The GIRAFFE spectra were reduced, extracted, and wavelength calibrated using the ESO pipeline\footnote{ftp://ftp.eso.org/pub/dfs/pipelines/giraffe/giraf-pipeline-manual-2.14.pdf}. Observations taken on the same date with the same MEDUSA plate were co-added  before extraction with the OPTIMAL method provided by the pipeline. Only one set of spectra was taken on 2012-Aug-21, with the MED1 plate. This was reduced separately via the SUM method, also provided by the pipeline, since the signal was too weak for the optimal extraction to work adequately. 

\begin{figure}
   \centering
   \includegraphics[width=\hsize]{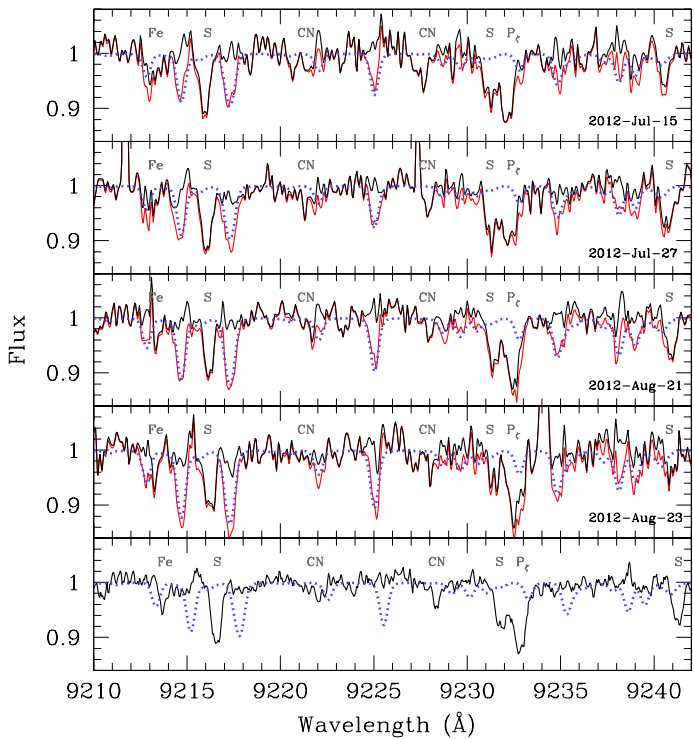}
      \caption{Top four panels show spectra of the star ET0048, for the dates of observation, in the rest frame of the Earth, before (red) and after (black) telluric correction. Bottom panel shows the final telluric-corrected, co-added spectrum, in the rest frame of the Sun. Blue dotted lines show the synthetic telluric spectra.
      }
         \label{fig:tellu}
   \end{figure}

The final reduced sets of spectra were sky-subtracted using a routine written by M. Irwin (also used by \citealt{Battaglia2008b}; Hill et al. in prep.), which scales the sky background to be subtracted from each object spectrum to match the observed sky emission lines. Correction for telluric absorption was done using synthetic telluric spectra from TAPAS\footnote{http://ether.ipsl.jussieu.fr/tapas/} \citep{Bertaux2014}, scaled to match the observed telluric lines in each set of stellar spectra. An example of the spectra, before and after telluric correction, is shown in Fig.~\ref{fig:tellu}.

Since the sets of spectra were combined from different numbers of observations, for each star they were co-added using the weighted median of the counts going into the spectra. The signal-to-noise  (S/N) ratios of the final co-added spectra were evaluated as the mean value over the standard deviation of the continuum in line-free regions. The average ratio of the sample is $\text{<S/N>}=62$ with a dispersion of $\sigma=12$. Table \ref{tab:main} lists the targets with positions, their velocity measurements and S/N ratios.

   \begin{figure}
   \centering
   \includegraphics[width=\hsize-0.4cm]{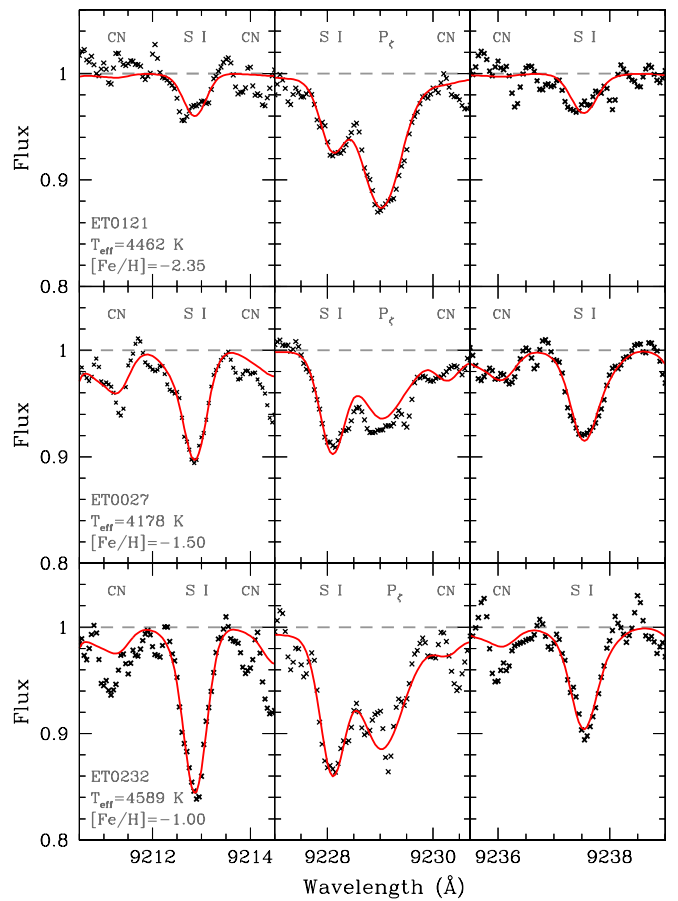}
      \caption{Spectra around the observed sulphur lines in three stars (ET0121, ET0027, ET0232) as (black) crosses. Solid (red) lines show the best fits to each line, as listed in Table~\ref{tab:abulines}. These three stars cover the metallicity range of the sample ($\text{[Fe/H]}=-2.35,-1.50,-1.00$), and have typical S/N ratios for the target stars, within 1$\sigma$ from the mean ($\text{S/N}=68,69,54$).
      }
         \label{fig:spectra}
   \end{figure}

%
\section{Stellar parameters and model atmospheres}

The stellar parameters (\Teff, \logg, and \vt), [Fe/H], and [Mg/H] of the target stars were determined by \hill\, and \citet{Asa2015}, and are listed in Table~\ref{tab:main}.

The  analysis was carried out using the spectral synthesis code TURBOSPEC\footnote{ascl.net/1205.004} developed by Bernand Plez \citep{AlvarezPlez1998,Plez2012}. The stellar atmosphere models are adopted from MARCS\footnote{marcs.astro.uu.se} \citep{Gustafsson2008} for stars with standard composition, 1D and assuming LTE, interpolated to match the exact stellar parameters for the target stars. Atomic parameters are adopted from the DREAM data base \citep{Biemont1999}, extracted via VALD\footnote{http://vald.astro.uu.se} (\citealt{Kupka1999} and references therein). The CN molecular parameters are from T. Masseron (private communication), derived with similar methods and laboratory data to those in  \citet{Brooke2014} and \citet{Sneden2014}.

Following \citet{Spite2011}, the adopted solar abundance is $\log\epsilon(\text{S})_\odot=7.16$ \citep{Caffau2011}, where we use the standard notation $\log\epsilon(\text{X})=\log(N_\text{X}/N_\text{H})+12$. Other solar abundances used here are $\log\epsilon(\text{Fe})_\odot=7.50$ and $\log\epsilon(\text{Mg})_\odot=7.58$, adopted from \citet{GrevesseSauval1998}. Literature data used in this paper are scaled to match these solar abundances.

\begin{table}
\caption{Atomic data for the measured S~I lines, adopted from the VALD database.}
\label{tab:sul}
\centering
\begin{tabular}{c c c c c}
\hline\hline
Wavelength      &               &       Transition      &       $\log{gf}$      &       $\chi_{ex}$     \\
(\AA)           &               &                               &                               &       (eV)            \\
\hline
9212.863        &       4s-4p           &       $^{5}$S$^{0}_{2} - ^{5}$P$_{3}$         &       0.42    &       6.525   \\
9228.093        &       4s-4p           &       $^{5}$S$^{0}_{2} - ^{5}$P$_{2}$         &       0.26    &       6.525   \\
9237.538        &       4s-4p           &       $^{5}$S$^{0}_{2} - ^{5}$P$_{1}$         &       0.04    &       6.525   \\

\hline
\end{tabular}
\end{table}

%
\section{Abundance measurements}

The sulphur abundances were determined from the high excitation multiplet 1 of S~I, at 9213, 9228, and 9238~\AA. The atomic data for these lines are listed in Table~\ref{tab:sul}. The central line, at 9228~\AA, is located in the wing of the Paschen~$\zeta$ line, which is taken into account with the synthetic spectra. In these relatively cold RGB stars ($T_\textsl{eff}\lesssim 4700$~K), most of the observed wavelength range is covered with CN molecular lines, which were used to estimate the [C/Fe] ratios in these stars (for more detail see \citealt{Asa2015}). An example of spectra around the measured sulphur lines is shown in Fig.~\ref{fig:spectra} for three target stars of typical S/N ratios for the sample.

Sulphur abundances of the first two lines (at 9213 and 9228~\AA) could be measured for all stars with the VLT/FLAMES spectra.  The weakest S line, at 9238~\AA, could not be measured in four of the more metal-poor stars ($\text{[Fe/H]}\lesssim-2$), and in those cases the abundance was determined from the other two lines. The abundance measurements for individual sulphur lines with their respective errors are listed in Table~\ref{tab:abulines}. 

\begin{figure}
   \centering
   \includegraphics[width=\hsize]{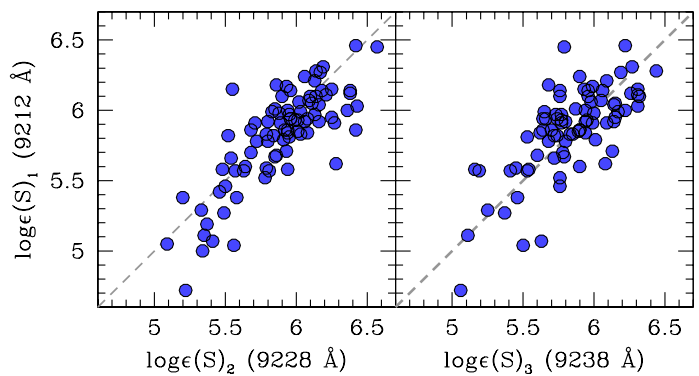}
      \caption{Relative LTE abundance measurements for the three sulphur lines for all the target stars.
      }
         \label{fig:S123}
   \end{figure}

The abundances from the three different sulphur lines are generally in good agreement with each other, as is shown in Fig.~\ref{fig:S123}. The average dispersion between lines in the sample of stars with three measured lines is $\text{<}\sigma\text{>}=0.14$, and the average difference between lines is $\text{<}\log\epsilon\text{(S)}_1-\log\epsilon\text{(S)}_2)\text{>}=-0.07$ and $\text{<}\log\epsilon\text{(S)}_1-\log\epsilon\text{(S)}_3)\text{>}=0.02$, with a dispersion of 0.21 and 0.23, respectively.

\subsection{Errors}

The error for individual sulphur lines is determined from the $\chi^2$ fit. The upper and lower error bars are defined as the moment the $\chi^2$ reaches a certain deviation from the best fit
\begin{equation}
\chi^2_\textsl{err}=(1+f)\chi^2_\textsl{bf}
\end{equation}
where $\chi^2_\textsl{bf}$ is the best fit and the constant factor $f$ is calibrated over the sample so that the average error is equal to the average dispersion between lines, $\text{<}\delta_\textsl{noise}\text{>}=\text{<}\sigma\text{>}=0.14$. The final error of a line, $\delta_\textsl{noise,i}$, is taken as the maximum value of the upper and lower error bars. 

In some cases the weakest line, at 9238~\AA, had a bigger error than the other lines, while in other cases the blending with the P$_\zeta$ or CN~lines, resulted in some lines being less reliable. To account for these effects, the abundance measurements of different S lines, $\log \epsilon \text{(S)}_i$, are weighted with their errors, $w_i$,  as 
\begin{equation}
\log \epsilon \text{(S)} = \frac{\sum\limits_{i=1}^{N_\text{S}} \log \epsilon (\text{S})_i \cdot w_i}{\sum\limits_{i=1}^{N_\text{S}} w_i}
\end{equation} 
where the sum goes over the number of S lines included. In most cases $N_\text{S}=3$, except the few cases where the line at 9238~\AA\ could not be measured. The weights of individual lines are defined as
\begin{equation}
w_i=\frac{1}{\delta^2_\textsl{noise,i}}
\end{equation}
where $\delta_\textsl{noise,i}$ is the statistical uncertainty of the abundance measurement of line $i$.\\

The final error is calculated as follows:

\begin{equation}
\delta_{\textsl{noise}} = \sqrt{\frac{N_S}{\sum^{N_S}_i w_i}}
\end{equation}

The systematic errors coming from the uncertainties of the stellar parameters $T_\textsl{eff}$, $\log g$, and $v_t$ are measured for abundance ratios $\Delta$[S/Fe]$_\text{sp}$ and $\Delta$[S/Mg]$_\text{sp}$, assuming $\Delta T_\textsl{eff}=100$~K, $\Delta \log g = 0.3,$ and $\Delta v_t=0.3$, as determined by Hill et al. in prep.\footnote{This is with the exception of the star ET0097, which has $\Delta T_\textsl{eff}=35$~K, $\Delta \log g = 0.13$, and $\Delta v_t=0.20$ \citep{Asa2015}.} They are added quadratically to the measurement error, $\delta_{\textsl{noise}}$, to get the adopted error:
\begin{equation}\label{eq:totalerr}
\delta_{\text{[S/X]}}=\sqrt{ \delta_{\textsl{noise}}(\text{S})^2+\delta_{\textsl{noise}}(\text{X})^2+ \Delta\text{[S/X]}_\text{sp}^2}
\end{equation}

%
\section{NLTE corrections}

The S~I lines of the high excitation multiplet 1, which are used here for the abundance determination, have previously been shown to be sensitive to NLTE effects \citep{Takeda2005,Korotin2008}. An extensive grid of NLTE corrections for F, G, and K stars was calculated in \citet{Takeda2005}. However, the target stars selected here are all high up on the RGB, and go lower in both \Teff\ and \logg\ than in the grid provided by \citet{Takeda2005}. Therefore, a new grid of NLTE corrections was calculated to fit the stellar parameters in our sample and is provided in Table~\ref{tab:nlte}.

The NLTE atomic model for sulphur includes 64 lower singlet, triplet, and quintet systems of S~I and the ground level of S~II \citep{Korotin2008,Korotin2009}. Oscillator strengths are adopted from TOP-base\footnote{http://cdsweb.u-strasbg.fr/topbase/topbase.html}, and photoionization cross sections are calculated assuming a hydrogen-like structure. The LTE stellar atmospheric models are from MARCS, and are used for the S abundance measurements of the target stars.

   \begin{figure} 
   \centering
   \includegraphics[width=\hsize-0.cm]{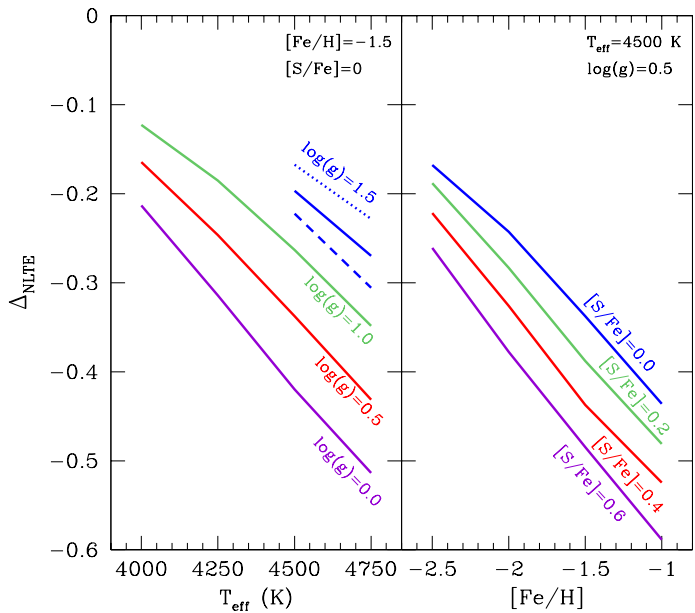}
      \caption{\textit{Left panel:} Solid lines show NLTE corrections for the S~I line at 9228~\AA\ as a function of \Teff\ for different \logg, assuming $\text{[Fe/H]}=-1.5$ and $\text{[S/Fe]}=0$. For $\log g=1.5$ the corrections for 9212~\AA\ are shown with a dashed line and with a dotted line for 9238~\AA. \textit{Right panel:} NLTE corrections for the sulphur line at 9228~\AA\ as a function of [Fe/H] for different [S/Fe], assuming $T_\textsl{eff}=4500$~K and $\log g=0.5$.
      }
         \label{fig:nlte}
   \end{figure}

      \begin{figure} 
   \centering
   \includegraphics[width=\hsize-0cm]{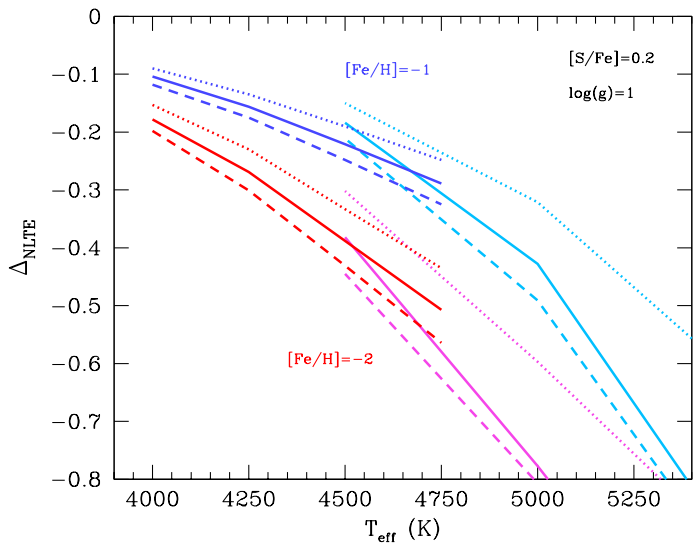}
      \caption{Comparison of the $h=0$, NLTE calculations of \citet{Takeda2005} and this work. The results from \citet{Takeda2005} for $T_\textsl{eff}\geq 4500$~K are shown in pink and light blue, while the results from this work are shown in red and blue for $T_\textsl{eff}\leq 4750$~K. The result is shown both for the smaller correction of $\text{[Fe/H]}=-1$ (blue and light blue) and the larger effect of $\text{[Fe/H]}=-2$ (red and pink). The calculations for all three S~lines are shown, 9213~\AA\ (dashed lines), 9228~\AA\ (solid lines), and 9238~\AA\ (dotted lines). In all cases $\log g = 1$ and $\text{[S/Fe]}=0.2$ are assumed. We note that \citet{Takeda2005} assumed a solar value of $\log \epsilon \text{(S)}_\odot=7.21$, while here a value of 7.16 is adopted, so there is a difference of 0.05 in [S/Fe]. In the case of \citet{Takeda2005}, $v_t=2$~km/s is assumed, but $v_t=1.7$~km/s in this work. These small discrepancies do not have a significant effect on the comparison.
      }
         \label{fig:takand}
   \end{figure}

As carefully described in \citet{Takeda2005}, the NLTE effects of the S~I lines at 9213, 9228, and 9238~\AA\ generally act in the direction of strengthening the lines, i.e. the correction is negative, $\Delta_\text{NLTE}=\text{[S/H]}_\text{NLTE}-\text{[S/H]}_\text{LTE}<0$. The dependence of the NLTE correction on \Teff\ and \logg\ is shown in Fig.~\ref{fig:nlte}. The NLTE effect clearly becomes stronger for lower gravity, higher temperature, and higher metallicity. However, the hotter stars tend to have higher $\log g$, so the trends of the NLTE corrections in the sample are not as strong as shown in Fig~\ref{fig:nlte}. The average NLTE correction for the target stars is $\text{<}\Delta_\text{NLTE}\text{>}=-0.29$ with a dispersion of $\sigma=0.05$. The NLTE corrections, $\Delta_\text{NLTE}$, for individual stars are listed in Table~\ref{tab:main}, and for individual lines of each star in Table~\ref{tab:abulines}.

The NLTE corrections are sensitive to the strength of the S~lines, which is here parameterized with [Fe/H] and [S/Fe]. The grid covers $\text{[Fe/H}]=-2.5,-2,-1.5,-1$, assuming LTE values: $\text{[S/Fe]}=0.0,0.2,0.4,0.6$. The temperature range is 
$T_\textsl{eff}=4000,4250,4500,4750$~K, and the entire grid is calculated for $\log g=0,0.5,1$. One exception to this is the case where $\log g =1$, $\text{[Fe/H]}=-2.5$, and $T_\textsl{eff}=4000$~K, which is not included. Finally, NLTE corrections for \logg=1.5 were calculated for the hottest temperatures, 4500 and 4750~K. The value $v_t=1.7$~km/s is adopted for the entire grid since the NLTE effects are not very sensitive to the turbulence velocity. Assuming $v_t=2.5$~km/s instead will make a $\lesssim 0.04$~dex change in the correction, which is much smaller than the measurement errors. Close to 90\% of the target stars have $1.4$~km/s~$\leq v_t \leq 2$~km/s, and  only one star is more than 0.7~km/s from the assumed value.

   \begin{figure}
   \centering
   \includegraphics[width=\hsize-0cm]{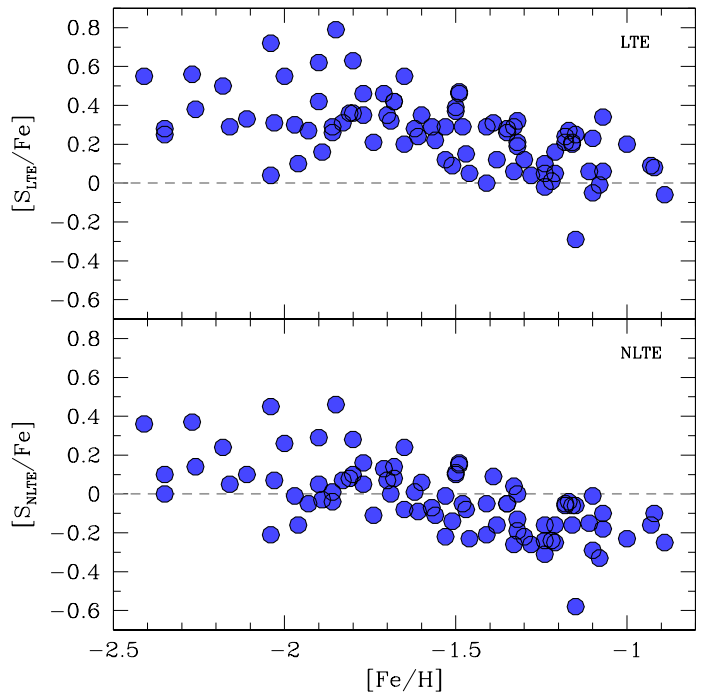}
      \caption{[S/Fe] vs. [Fe/H] for the target stars. \textit{Upper panel:} Assuming LTE. \textit{Lower panel:} Including NLTE corrections.
      }
         \label{fig:FeLTE}
   \end{figure}

The grid of NLTE corrections provided by \citet{Takeda2005} goes down to $T_\textsl{eff}=4500$~K, and $\log g =1.0$, so it is possible to make a comparison with their work, see Fig~\ref{fig:takand}. The two different calculations show comparable results over the small overlapping region of stellar parameters.\footnote{For \cite{Takeda2005}, we plot the fiducial case of $h=0$, where $h$ is the logarithm of the H~I correction factor applied to the classical formula (see their paper for details).}

  \begin{figure*}
   \centering
   \includegraphics[width=\hsize-2cm]{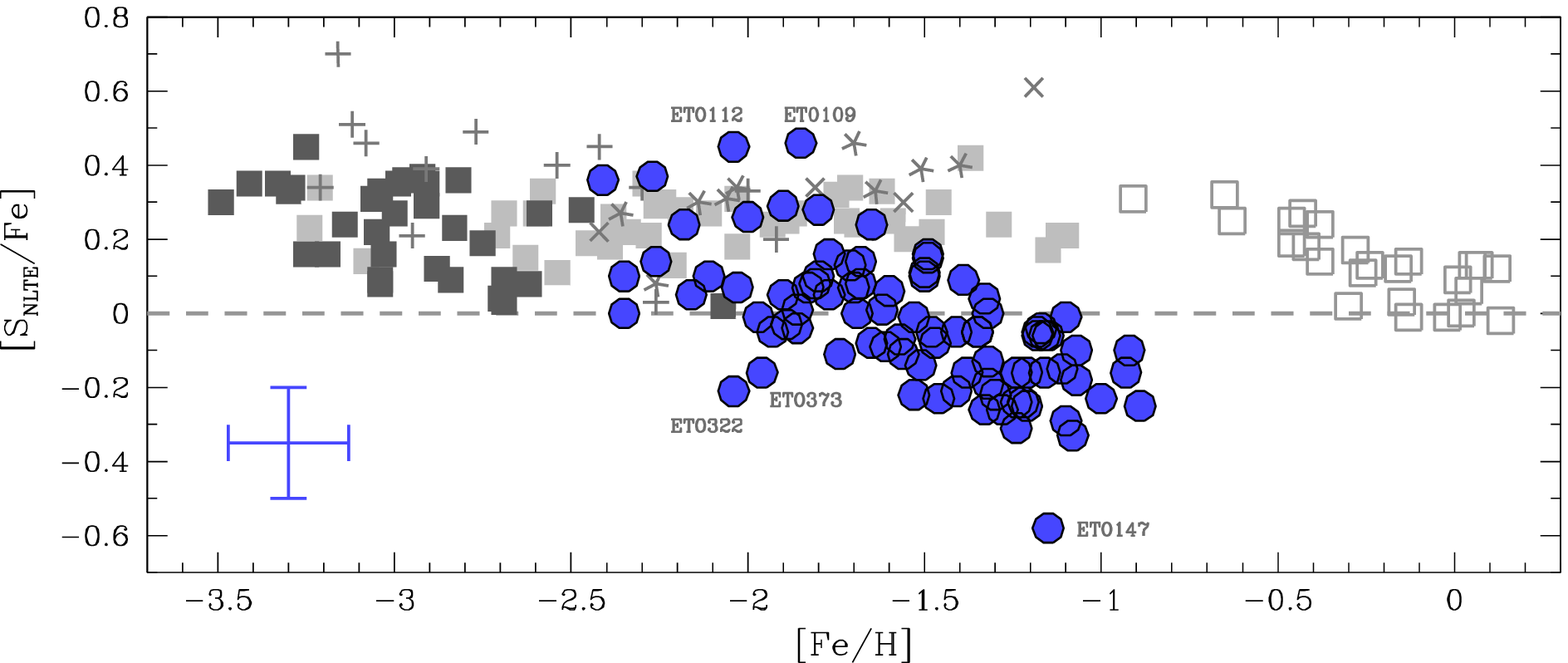}
      \caption{NLTE-corrected values of [S/Fe] for the target stars in Sculptor (blue circles). Filled squares show NLTE sulphur measurements from multiplet 1 in Galactic halo stars taken from \citet{Nissen2007} (light grey) and \citet{Spite2011} (dark grey). Sulphur abundances in halo stars determined from multiplet 3 are shown with pluses \citep{TakedaTakada-Hidai2012}, asterisks \citep{Jonsson2011}, and exes \citep{Caffau2010}. In all cases, NLTE corrections have been applied. Milky Way disc stars from \citet{Chen2002} are shown with open grey squares. The disc stars have not been corrected for NLTE effects, but these sulphur abundances are measured with lines that are not believed to be sensitive to NLTE effects, so any corrections applied would be small (and negative). No 3D corrections have been applied to any of the data, but are expected to be positive for both multiplet 1 and 3. A representative error bar for the Sculptor data is shown. 
      }
         \label{fig:FeNLTE}
   \end{figure*}

%
\section{Results and discussion}

The final sulphur abundances with NLTE corrections are provided in Table~\ref{tab:main}. The measured values of [S/Fe], under the assumptions of LTE and NLTE, are shown in Fig.~\ref{fig:FeLTE}. The negative correction lowers the measured abundances, and slightly reduces the scatter, from $\sigma=0.19$ to 0.16 at $\text{[Fe/H]}\leq-2$ and from $\sigma=0.15$ to 0.14, both at $-2<\text{[Fe/H]}\leq-1.5$ and $-1.5<\text{[Fe/H]}$. A comparison between the NLTE-corrected sulphur abundances in Sculptor and the Galactic halo is shown in Fig.~\ref{fig:FeNLTE}.

\subsection{Possible outliers}
The scatter of [S/Fe] measurements in Sculptor seems to increase below $\text{[Fe/H]}\approx-1.8$, see Fig.~\ref{fig:FeNLTE}, where $\sigma=0.17$, slightly exceeding what is expected from the typical error in this region, $\text{<}\delta_\text{[S/Fe]}\text{>}=0.15$. It is important to take into account that at these low metallicities, the sulphur lines are often very weak, so for three of the stars below $\text{[Fe/H]}\leq-2.2$, the S~abundance is determined from only two lines, and the other two stars in this metallicity range show significant scatter between the three lines, $\sigma\approx0.27$.

However,  two stars at $\text{[Fe/H]}\approx-2$, ET0112 and ET0109, show [S/Fe] of about $\sim0.65$~dex higher than two S-poor stars at similar metallicities, ET0322 and ET0373, see Fig.~\ref{fig:FeNLTE}. The S-rich stars, ET0112 and ET0109, both have high S/N ratios, more than $1\sigma$ above the mean of the sample, and are consistent (within the errors) with the trend seen in the Galactic halo. The S-poor stars, ET0322 and ET0373, have S/N lower than average, but within $1\sigma$ of the mean.  Of these, ET0373 only has two measurable S lines, but has a normal value of sulphur when compared to magnesium, $\text{[S/Mg]}~=~-0.50$, and could therefore be slightly low in $\alpha$-elements in general. However, ET0322 shows the very low value of $\text{[S/Mg]}=-0.95$, and seems to be an outlier in sulphur. If that one star is ignored, the scatter below $\text{[Fe/H]}\leq-1.8$ is consistent with the errors.

      \begin{figure}
   \centering
      \includegraphics[width=\hsize-1cm]{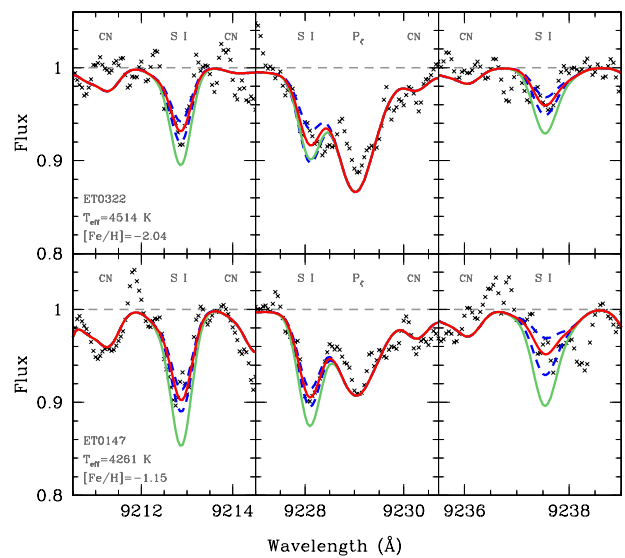}
      \caption{Spectra for the S-low stars, ET0322 and ET0147, are shown with black crosses. Red solid lines are the best fits of each S~line, and blue dashed lines are the measurement errors, $\delta_\textsl{noise}$, as listed in Table~\ref{tab:abulines}. Green solid lines show [S/Fe]=0.40 in the upper panel, and [S/Fe]=0.15 in the lower panel, which are the average (LTE) S~abundances in the sample, in the range $\pm0.25$ from the [Fe/H] of each star.
      }
         \label{fig:Slow}
   \end{figure}

Another clear outlier is seen at a higher metallicity, ET0147 at $\text{[Fe/H]}=-1.15$, which is also low in $\text{[S/Mg]}$. Neither of these stars that seem low in sulphur, ET0322 or ET0147, stand out  in measurements of other elements (such as Mg, Ca, Ti, Cr, Ba, and La) when compared to other stars in Sculptor (Hill et al. in prep.). Both ET0322 and ET0147 have reasonable S/N ratios, within $1\sigma$ of the mean value, and we have checked that these low S values are not being driven by poor placement of continuum, and ensured that the error estimates,  $\approx0.2$~dex in both cases, are reasonable. We do note, however, that for the star ET0147, there is a difference of 160~K in $T_\textsl{eff}$ as determined by Hill et al. in prep, which is used here, and by photometry from \citet{deBoer2011} following the recipe from \citet{RamirezMelendez2005}. This is the largest deviation in the sample, and it is therefore possible that we are underestimating the errors coming from the stellar parameters for this particular star. For ET0322, the difference in $T_\textsl{eff}$ between these two methods is well within the errors, and similar to other stars in the sample.

The spectra for these possible outliers, ET0322 and ET0147 are shown in Fig.~\ref{fig:Slow}. In the case of ET0322, the central line at 9228~\AA\   is consistent with the average value of [S/Fe] in the sample of stars with similar [Fe/H]. However, in this case the blending with the Paschen~$\zeta$ line is not perfectly reproduced by the synthetic spectra, and causes it to be the least reliable  of the three available lines. The other two lines are not consistent with the normal trend. For ET0147, Fig.~\ref{fig:Slow} shows that this star is not consistent with the average [S/Fe] of the sample with similar [Fe/H] unless we are severely underestimating the errors of the stellar parameters. There is, however, no obvious scenario where stars are expected to be low in sulphur and not in the other $\alpha$-elements, so these outliers should not be interpreted too strictly, especially without further confirmation since, considering the errors, they are still consistent with the range of scatter.

\subsection{The trend of [S/Fe] with [Fe/H]}

The measurements of [S/Fe] in the Galactic halo show a plateau at low metallicities, both when measured with multiplet 1 at 9210-9240 \AA, and multiplet 3 at 10455-10459~\AA, see Fig.~\ref{fig:FeNLTE}. The level of this plateau is, however, higher when measured with multiplet 3 compared to multiplet 1. The statistical errors of the mean can be defined as $\delta_\text{SE}=\sigma / \sqrt{N-1}$, where $N$ is the number of stars in each sample. Therefore, we get $\text{<[S/Fe]}\text{>}_\text{mul1}=0.22$ with 
$\sigma=0.14$ and $\delta_\text{SE}=0.01$, while $\text{<[S/Fe]}\text{>}_\text{mul3}=0.36$  with $\sigma=0.09$ and $\delta_\text{SE}=0.03$. Even when the two possible outliers, high in S, at $\text{[Fe/H]}=-1.19$ \citep{Caffau2010} and $\text{[Fe/H]}=-3.16$ \citep{TakedaTakada-Hidai2012} are excluded, we get $\text{<[S/Fe]}\text{>}_\text{mul3}=0.33$ with $\sigma=0.12$ and $\delta_\text{SE}=0.02$.

The reason for this discrepancy is unknown, but possibly arises from the NLTE corrections or a difference in 3D corrections between the two multiplets, which have not been applied here. The 3D correction for multiplet 1 is expected to be small and positive, $\lesssim+0.1$ \citep{Nissen2004,Spite2011}. For multiplet 3, however, the 3D corrections are expected to be larger, $\sim+0.2$ \citep{Jonsson2011}, making the discrepancy between the results from the two multiplets even larger. An extensive sample of sulphur measurements from both multiplets in the same stars would be helpful to explore this issue further and determine the exact level of the plateau. Here, however, our main interest lies in comparing the Sculptor results with the Galactic halo, so in order to avoid this issue, we will limit our comparison to the results of \citet{Nissen2007} and \citet{Spite2011}, which use the same S lines as the results obtained here.
We note that \citet{Spite2011} use the same atomic model for the NLTE corrections as is used here \citep{Korotin2008}, while \citet{Nissen2007} use the NLTE calculations from \citet{Takeda2005}.

\begin{figure}
   \centering
   \includegraphics[width=\hsize]{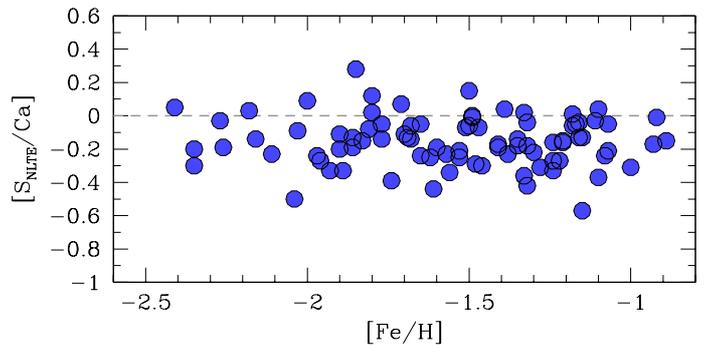}
      \caption{[S/Ca] vs. [Fe/H] for the target stars.
      }
         \label{fig:SCa}
   \end{figure}

The chemical enrichment of the Milky Way was apparently a much more rapid process than in Sculptor, the small dwarf spheroidal galaxy, where star formation was inefficient. During the first 1-2~Gyr of star formation, before Type~Ia~SNe started to contribute, the Galactic halo was enriched to much higher metallicities ($\text{[Fe/H]}\sim-1$) than Sculptor ($\text{[Fe/H]}\sim-1.8$). Therefore, the `knee' of [S/Fe], where this abundance ratio starts to decrease as a result of a growing contribution from SNe~Type~Ia to [Fe/H], is seen at a higher metallicity in the Galactic halo, where S also shows similar behaviour to other $\alpha$-elements, e.g. Si and Ca \citep{Chen2002,Cayrel2004}. The position of the knee of [S/Fe] in Sculptor is comparable to what is previously measured in the other $\alpha$-elements in this galaxy (\citealt{Tolstoy2009}; Hill et al. in prep); as is shown for example in Fig.~\ref{fig:SCa}, S follows the $\alpha$-element Ca over the observed iron range.

If a plateau-like behaviour of [S/Fe] in Sculptor is assumed at lower metallicities, $\text{[Fe/H]}\leq-2$, the plateau lies at $\text{<[S/Fe]}_\text{Scl}\text{>}=0.16$, with $\sigma=0.18$ and $\delta_\text{SE}=0.06$. If the outlier ET0322 is not included, we get the higher value of $\text{<[S/Fe]}_\text{Scl}\text{>}~=~0.19$ with $\sigma=0.15$ and $\delta_\text{SE}=0.05$. In the Galactic halo, \citet{Nissen2007} found an average value of $\text{<[S/Fe]}_\text{halo,N}\text{>}~=~0.21$ with $\sigma=0.07$ $\delta_\text{SE}=0.01$, and   in the sample of \citet{Spite2011}, $\text{<[S/Fe]}_\text{halo,S}\text{>}~=~0.23\pm0.12$ with $\sigma=0.12$ $\delta_\text{SE}=0.02$. Adopting $\delta_\text{SE}$ as the error of the mean, all three surveys are in agreement (even when the possible outlier in Sculptor is included). Although in this case the errors are underestimated since we are not taking into account any possible systematic errors of the samples such as the determination of stellar parameters or possible differences in the NLTE corrections between surveys. 

The Sculptor sample is therefore consistent with having experienced the same early chemical enrichment of sulphur as observed in the Galactic halo, although S measurements of stars with $\text{[Fe/H]}\leq-2.5$ are needed to further confirm this.

      \begin{figure}
   \centering
   \includegraphics[width=\hsize]{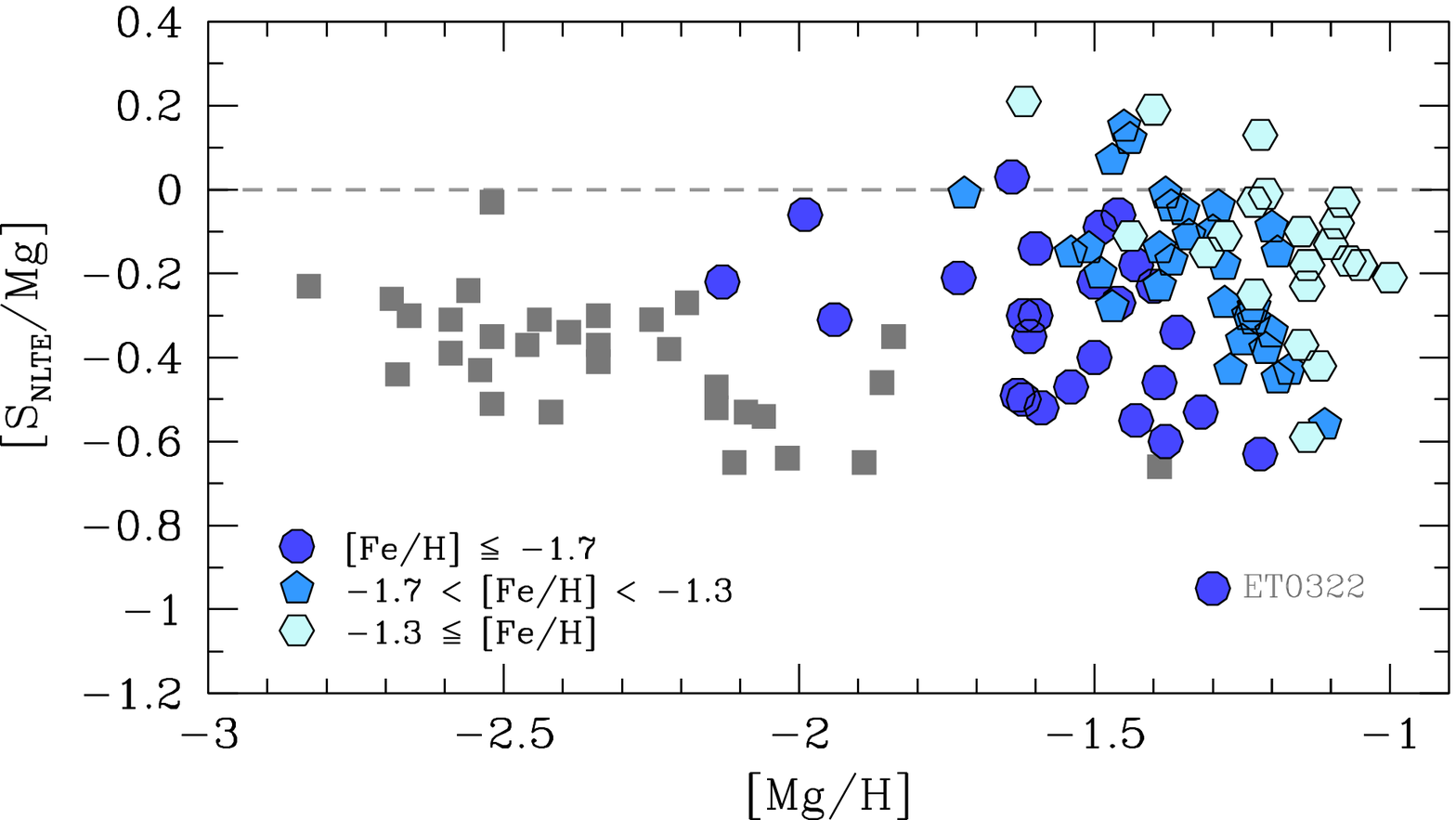} 
      \caption{Abundance ratios [S$_\text{NLTE}$/Mg] with [Mg/H]. Grey squares show Galactic halo stars with S measurements from \citet{Spite2011}, and Mg from \citet{Andrievsky2010}. Circles, pentagons, and hexagons show the target stars in Sculptor; the most iron-poor stars ($\text{[Fe/H]}\leq-1.7$) as dark blue circles, the most iron-rich ($-1.3\leq\text{[Fe/H]}$) are shown as cyan hexagons, and other stars as blue pentagons. Note that in Sculptor, the Mg abundances have not been corrected for NLTE effects. However, these are expected to be small (and positive) for the target stars ($\lesssim 0.1$~dex).
      }
         \label{fig:Mg}
   \end{figure}

\subsection{Comparison of [S/Fe] and [S/Mg]}
It has been noted by \citet{Spite2011} that the scatter of S abundance ratios in the Galactic halo is slightly larger when Mg is used as a reference element instead of Fe. This is not expected since Mg and S are believed to form in similar processes, normally the hydrostatic burning of C, O, and Ne, while Fe is more affected by explosive nucleosynthesis, mixing, and fallback.

The S abundances of our target stars are shown with respect to Mg in Fig.~\ref{fig:Mg}, and the scatter is clearly bigger than when compared to Fe. In addition, the average level of [S/Mg] is higher over the entire Sculptor sample, $\text{<} \text{[S/Mg]}_\text{Scl}\text{>}=-0.23$ with $\delta_\text{SE}=0.02$, compared to the Galactic halo, $\text{<} \text{[S/Mg]}_\text{halo,S}\text{>}=-0.39$ with $\delta_\text{SE}=0.03$.

It should be noted that the Mg abundances in Sculptor have not been corrected for NLTE effects as has been done for the \citet{Spite2011} sample. As shown in \citet{Andrievsky2010}, this correction is expected to be positive thus lowering the [S/Mg] ratio, but this effect is weaker at higher metallicities and in colder stars (the average \Teff\ of the Sculptor stars is $\approx4300$~K, while it is  $\approx4900$~K for the giants in \citealt{Spite2011}), so these corrections are expected to be small ($\lesssim 0.1$~dex) in the Sculptor sample.

\begin{figure}
   \centering
   \includegraphics[width=\hsize]{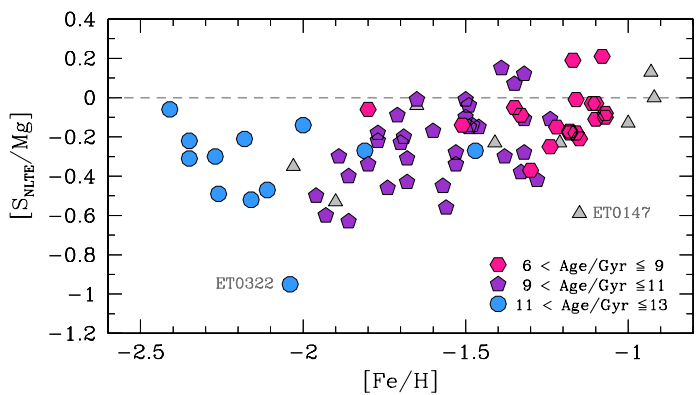}
      \caption{Results of [S$_\text{NLTE}$/Mg] versus [Fe/H] for the Sculptor sample. Ages are shown with colours, as estimated by \citet{deBoer2012}, while grey triangles show stars with no age estimate. The average error on the ages is 1.8~Gyr.
      }
         \label{fig:SMgFe}
   \end{figure}

Another possible reason for the shift in the average value of [S/Mg] in Sculptor compared to the halo is that [S/Mg] shows a slightly increasing trend with [Fe/H], see Fig.~\ref{fig:SMgFe}, where there is also an age gradient. Although Mg and S form in similar processes, it is possible that they have slightly different production chains \citep{LimongiChieffi2003}; this, however, does not explain the trend seen with age and metallicity. The observed slope is not created by the lack of NLTE corrections for Mg; those corrections would create a slope in the opposite direction, since the correction is higher at lower metallicities. 

Therefore, this trend of [S/Mg] as [Fe/H] increases, most probably comes from an increasing SN~Ia contribution. Although SN~Ia yields have very high ratios of iron-peak elements, they also produce a non-negligible amount of some of the $\alpha$-elements (such as Si, S, and Ca). In particular, the [S/Mg] ratio in Type~Ia~SNe yields is more than an order of magnitude higher than predicted for SN~II \citep{Iwamoto1999}. This can explain the increasing [S/Mg] ratio as the SN~Ia contribution becomes more important. In fact, a similar trend of [Ca/Mg] with [Fe/H] is seen in the data of Hill et al. in prep. that is consistent with this explanation. 

To make a meaningful comparison with the Galactic halo sample, we therefore only use stars with $\text{[Fe/H]}~\leq~-2$, where the contribution from Type~Ia SNe is thought to be minimal. This gives a mean of $\text{<[S/Mg]}_\text{Scl}\text{>}~=~-0.37$, with $\sigma=0.25$ and $\delta_\text{SE}=0.08$. If the star ET0322, with $\text{[S$_\text{NLTE}$/Mg]}~=~-0.95$, is excluded the result is $\text{<[S/Mg]}_\text{Scl}\text{>}~=~-0.30$, with $\sigma=0.16$ and $\delta_\text{SE}=0.06$. This is in agreement with the value observed in the Galactic halo, $\text{<[S/Mg]}_\text{Halo}\text{>}~=~-0.39$, with $\sigma=0.15$ $\delta_\text{SE}=0.03$ \citep{Spite2011}. If the average NLTE correction of Mg in Sculptor is 0.2~dex or more, these results would no longer be in agreement, which supports the assumption that only a modest correction is needed for our sample stars.

As was seen in the \citet{Spite2011} sample for the Galactic halo, in Sculptor stars with $\text{[Fe/H]}\leq-2$, the scatter of [S/Mg]$_\text{Scl}$ ($\sigma=0.25$) is larger than for [S/Fe]$_\text{Scl}$ ($\sigma=0.18$). However, the measurement errors of [Mg/H] are bigger on average than for [Fe/H] in the Sculptor sample, giving the final average error $\text{<}\delta_\text{[S/Mg]}\text{>}=0.27$, while $\text{<}\delta_\text{[S/Fe]}\text{>}=0.16$. In our sample, it therefore seems likely that the scatter is dominated by measurement errors, rather than showing an intrinsic scatter, and more precise measurements are needed to clarify this issue.

%
\section{Conclusions}

From the DART survey, detailed abundance measurements are known for $\sim$100 RGB stars in Sculptor, covering the metallicity range $-2.5\leq\text{[Fe/H]}\leq-0.8$ (\citealt{Tolstoy2009}; Hill et al. in prep.). High-resolution VLT/GIRAFFE spectra, with grating HR22B, were acquired for 86 of those stars to measure the sulphur abundances from the triplet at 9213, 9228, and 9238~\AA. These sulphur lines have been shown to be sensitive to NLTE effects \citep{Takeda2005,Korotin2008}. Therefore, NLTE corrections have been calculated for a grid of stellar parameters covering the observed sample, and are available in Table \ref{tab:nlte}.

 Although the sample includes only 12 stars with $\text{[Fe/H]}\leq-2$, we find tentative evidence for a plateau in the ratio [S/Fe] at $\text{[Fe/H]}\leq-2$, similar to that seen for sulphur in the Milky Way halo \citep{Nissen2007,Spite2011} and for other $\alpha$-elements both in Sculptor (\citealt{Tolstoy2009}; Hill et al. in prep) and the halo (e.g. \citealt{Cayrel2004}). At the lowest metallicity end, $\text{[Fe/H]}\leq-2$, the scatter is rather large, $\sigma_\text{[S/Fe]}=0.18$, but if one possible outlier is excluded, it is consistent with measurement errors.  At higher metallicities, [S/Fe] decreases with increasing [Fe/H], reaching subsolar values at $\text{[Fe/H]}\gtrsim-1.5$, similar to what is seen in other $\alpha$-elements in Sculptor (\citealt{Tolstoy2009}; Hill et al. in prep.). This occurs at lower [Fe/H] than the halo, likely owing to a less efficient star formation.

Similarly to the sample of \citet{Spite2011}, the scatter in S is bigger when Mg is used as a reference element rather than Fe, contrary to theoretical predictions. However, for our results, this is consistent with the errors on the measurements of these elements. Our sample of Sculptor stars also shows an increase in [S/Mg] with [Fe/H], presumably due to an increasing SN~Ia contribution to [S/Mg].

\begin{acknowledgements}

The authors are indebted to the International Space Science Institute (ISSI), Bern, Switzerland, for supporting and funding the international team ``First stars in dwarf galaxies''. S.~A.~K. and S.~M.~A. acknowledge SCOPES grant No. IZ73Z0-152485 for financial support. S.~S. acknowledges support from the Netherlands Organisation for Scientific Research (NWO), VENI grant 639.041.233.
\end{acknowledgements}


\setcounter{table}{0}
\renewcommand{\thetable}{A.\arabic{table}}

\Online

\enlargethispage{20cm}
\nopagebreak
\begin{sidewaystable*}
\centering
\tiny
\tabcolsep=0.11cm

\begin{minipage}{\linewidth-2cm}
\centering%
\tabcaption{Positions and stellar parameters of the target stars in Sculptor, including Fe, Mg, and S abundance measurements and NLTE corrections for S. The complete table is available online.}%
\label{tab:main}
\begin{tabular}{lccccccccccccccccccc}
\hline\hline
Star    &       RA      &       Dec     &       $T_\textsl{eff}$        &       $\log g$      &       $v_t$   &       $v_r$   &       S/N     &       [Fe/H]  & $\delta_\text{[Fe/H]}$  &       [Mg/H]  &       
$\delta_\text{[Mg/H]}$  &       $N_S$   &       $\log \epsilon (\text{S})$      &       $\delta_\textsl{noise}$ &       $\Delta_\textsl{NLTE}$  &       [S/Fe]  &       
$\delta_\text{[S/Fe]}$  &       [S/Mg]  &       $\delta_\text{[S/Mg]}$  \\
        &       (J2000) &       (J2000) &       (K)     &               &       (km/s)  &       (km/s)  &               &               &               &               &               &               &       (LTE)   &               &               &       (NLTE)  &               &       (NLTE)  &       \\
\hline
ET0024  &       1  00  34.04    &       $-33$  39  04.6 &       3897    &       0.0     &       2.2     &       113.3   &       94      &$      -1.24   $&      0.10    &$      ...     $&      ...     &       3       &       5.96    &       0.14    &$      -0.21   $&$     -0.16   $&      0.20    &$      ...     $&      ...     \\
ET0026  &       1  00  12.76    &       $-33$  41  16.0 &       4245    &       0.5     &       1.7     &       97.1    &       76      &$      -1.80   $&      0.16    &$      -1.36   $&      0.22    &       3       &       5.71    &       0.11    &$      -0.26   $&$     0.09    $&      0.12    &$      -0.35   $&      0.24    \\
ET0027  &       1  00  15.37    &       $-33$  39   06.2        &       4178    &       0.3     &       2.2     &       111.8   &       69      &$      -1.50   $&      0.13    &$      -1.38   $&      0.19    &       3       &       6.04    &       0.09    &$      -0.29   $&$     0.09    $&      0.12    &$      -0.03   $&      0.25    \\
ET0028  &       1  00  17.77    &       $-33$  35  59.7 &       4085    &       0.3     &       2.0     &       119.8   &       73      &$      -1.22   $&      0.11    &$      -1.31   $&      0.16    &       3       &       5.95    &       0.13    &$      -0.25   $&$     -0.24   $&      0.18    &$      -0.15   $&      0.26    \\
ET0031  &       1  00   07.57   &       $-33$  37   03.9        &       4329    &       0.5     &       2.1     &       113.3   &       72      &$      -1.68   $&      0.17    &$      -1.17   $&      0.22    &       3       &       5.90    &       0.10    &$      -0.33   $&$     0.08    $&      0.11    &$      -0.43   $&      0.23    \\
\hline
\end{tabular}

\end{minipage}

\end{sidewaystable*}

\clearpage

\begin{minipage}{\linewidth+\linewidth-1cm}
\centering%
\small
\tabcaption{Abundances for individual sulphur lines with their associated measurement errors and NLTE corrections. The complete table is available online.}%
\label{tab:abulines}%
\begin{tabular}{lccccccccc}
\hline\hline
Star    &$      \log\epsilon{(\text{S}_{1})}    $&$     \delta_{\textsl{noise,1}} $&$     \Delta_{\text{NLTE,1}}$&$       \log\epsilon{(\text{S}_{2})}    $&$     \delta_{\textsl{noise,2}}$&$             \Delta_{\text{NLTE,2}}          $&$     \log\epsilon{(\text{S}_{3})}    $&$\delta_{\textsl{noise,3}}            $&$ \Delta_{\text{NLTE,3}}          $\\
&(9213~\AA)&(9213~\AA)&(9213~\AA)&(9228~\AA)&(9228~\AA)&(9228~\AA)&(9238~\AA)&(9238~\AA)&(9238~\AA)\\
\hline
ET0024  &       5.93    &       0.14    &$      -0.22   $&      6.01    &       0.14    &$      -0.21   $&      5.95    &       0.14    &$      -0.19   $\\
ET0026  &       5.68    &       0.10    &$      -0.29   $&      5.86    &       0.12    &$      -0.26   $&      5.60    &       0.12    &$      -0.22   $\\
ET0027  &       5.92    &       0.08    &$      -0.33   $&      6.06    &       0.10    &$      -0.30   $&      6.14    &       0.08    &$      -0.25   $\\
ET0028  &       5.98    &       0.16    &$      -0.28   $&      5.88    &       0.12    &$      -0.25   $&      6.00    &       0.12    &$      -0.22   $\\
ET0031  &       5.93    &       0.10    &$      -0.38   $&      5.99    &       0.10    &$      -0.34   $&      5.77    &       0.10    &$      -0.28   $\\

\hline
&&&&&&&&&\\
&&&&&&&&&\\
&&&&&&&&&\\
\end{tabular}
\end{minipage}

\hspace{2cm}

\begin{minipage}{\linewidth+1.5cm}
\centering%
\tabcaption{Example of NLTE corrections for a grid of stellar parameters fitting the target stars, assuming $v_t=1.7$~km/s. The complete table is available online.}%
\label{tab:nlte}%
\begin{tabular}{cccccccc}
\hline\hline
$T_\textsl{eff}$        &       [Fe/H]  &       $\log g$        &       [S/Fe]  &       $\Delta_\text{NLTE}$    &       $\Delta_\text{NLTE}$    &       $\Delta_\text{NLTE}$    \\
(K)     &               &               &               &       (9213 \AA)      &       (9228 \AA)    &       (9238 \AA)      \\
\hline
4000    &$      -2.5    $&      0.0     &       0.0     &$      -0.125  $&$     -0.109  $&$     -0.089  $\\
4000    &$      -2.5    $&      0.5     &       0.0     &$      -0.033  $&$     -0.008  $&$     -0.008  $\\
4000    &$      -2.0    $&      0.0     &       0.0     &$      -0.151  $&$     -0.124  $&$     -0.089  $\\
4000    &$      -2.0    $&      0.5     &       0.0     &$      -0.130  $&$     -0.113  $&$     -0.092  $\\
4000    &$      -2.0    $&      1.0     &       0.0     &$      -0.099  $&$     -0.089  $&$     -0.077  $\\

\hline
\end{tabular}

\end{minipage}


\end{document}